\documentclass{jps-cp}
\usepackage{txfonts} 

\usepackage{bm}
\usepackage{graphicx}
\usepackage{ascmac}
\usepackage{amsmath,amssymb}
\usepackage{cancel}
\usepackage{braket}

\usepackage{color}

\title{Structure of near-threshold states in systems with Coulomb and short-range interactions}

\author{Tomona \textsc{Kinugawa}$^{1}$ and Tetsuo \textsc{Hyodo}$^{2}$}

\inst{$^{1}$Nishina Center for Accelerator-Based Science, RIKEN, Wako 351-0198, Japan \\
$^{2}$Research Center for Nuclear Physics, The University of Osaka, Ibaraki, Osaka 567-0047, Japan}

\email{tomona.kinugawa@riken.jp}

\abst{We study the nature of near-threshold eigenstates in systems with the attractive Coulomb plus short-range interactions. Using a model providing the Coulomb-modified effective range expansion, we analyze pole trajectories and the internal structure of near-threshold states characterized by the compositeness. We find that bound state and resonance poles are disconnected in the presence of the attractive Coulomb interaction, in contrast to the repulsive Coulomb force. Near the threshold, bound states become almost purely composite, while the behavior of the compositeness near unityis controlled by the competition between the Coulomb and short-range interactions, characterized by the Bohr radius and the Coulomb effective range.
}

\kword{Hadronic molecule, Compositeness, Clustering phenomena}

\begin{document}
\maketitle

\section{Introduction}

Understanding the internal structure of exotic hadrons is expected to provide insight into nonperturbative phenomena of the strong interaction. In contrast to ordinary hadrons, which are composed of a quark-antiquark pair (mesons) or three quarks (baryons), exotic hadrons can have more complex internal structures~\cite{Hosaka:2016pey,Guo:2017jvc,Brambilla:2019esw,Hosaka:2025gcl}. Their wave functions can be expressed as superpositions of various components with different structures. In addition to the ordinary structures, one possible component is a multiquark, which corresponds to a compact state composed of four or more quarks. Another one is the hadronic molecule, a weakly bound state of two hadrons in which each hadron retains its identity as a hadronic (color-singlet) degree of freedom. Experimentally, many exotic hadrons have been observed near thresholds, which correspond to the lowest energies at which scatterings occur. Well-known examples include the $X(3872)$~\cite{Belle:2003nnu} near the $D\bar{D}^{*}$ threshold and 
the $T_{cc}(3875)^{+}$~\cite{LHCb:2021auc,LHCb:2021vvq} near the $DD^{*}$ threshold. Therefore, it is necessary to understand the properties of near-threshold states for elucidating the internal structure of exotic hadrons.

The internal structure of near-threshold bound states can be quantified by a measure known as the compositeness~\cite{Weinberg:1965zz,Hyodo:2013nka,vanKolck:2022lqz,Kinugawa:2024crb}. The compositeness $X$ is defined as the overlap between the free scattering states $\ket{\bm{p}}$ and the bound state $\ket{B}$:
\begin{align}
X &= \int \frac{d^{3}p}{(2\pi)^{3}}|\braket{\bm{p}|B}|^{2}.
\label{eq:X-def}
\end{align}
In this sense, $X$ can be interpreted as the probability of finding a hadronic molecular component in the hadron wave function. Using the compositeness, the internal structure of a bound state can be characterized in terms of whether the hadronic molecular component is dominant or not. In a broader perspective, the compositeness serves as an indicator of the fraction of cluster components, not only for hadronic molecules with a quark-cluster structure but also for other composite systems. Therefore, the concept of compositeness can be applied not only to exotic hadrons but also to nuclear and atomic systems.

In the near-threshold energy region of the scatterings with the short-range interactions, the scattering length diverges, and physical observables are governed solely by the scattering length. As a result, systems with different microscopic structures are known to follow the same universal behavior. This is called the low-energy universality~\cite{Braaten:2004rn,Naidon:2016dpf}. To discuss the internal structure of near-threshold states, we therefore focus on this universality. As a consequence of the low-energy universality, shallow bound states located below the threshold in short-range interactions tend to be composite dominant, with the compositeness close to unity~\cite{Kinugawa:2022fzn,Kinugawa:2023fbf}. In contrast, near-threshold resonances that appear slightly above the threshold are shown to be non-composite-dominant~\cite{Hyodo:2013iga,Matuschek:2020gqe,Kinugawa:2024kwb}. In this way, the nature of near-threshold states is well understood from the viewpoint of the low-energy universality.

However, there also exist near-threshold states observed in two-body systems where not only short-range interactions but also the Coulomb interaction plays a role. A well-known example is provided by nuclear $\alpha$-cluster systems. The ${}^{8}\mathrm{Be}$ nucleus is a resonance of the 
two-$\alpha$ (${}^{4}\mathrm{He}$ nuclei) system~\cite{Braaten:2004rn,Higa:2008dn}. As an example in hadron physics, lattice QCD calculations report that, in the $\Omega_{ccc}$-$\Omega_{ccc}$ system, the dibaryon resonance appears slightly above the threshold~\cite{Lyu:2021qsh}. These two states are theoretically found to become bound states below the threshold if the repulsive Coulomb interaction is switched off and only the strong interaction is taken into account. In contrast, in the presence of the repulsive Coulomb interaction, they are pushed above the threshold and appear as resonances. A similar mechanism is known to occur also in systems with an attractive Coulomb interaction. For example, the bound state of $\Xi^{-}$ and $\alpha$, one of the $\Xi$ hypernuclei, is considered to be a Coulomb-assisted bound state, which would be unbound in the absence of the Coulomb attraction. In this way, although the Coulomb interaction is much weaker than the strong interaction, it is known to have a significant contribution on the properties of near-threshold states. Therefore, in near-threshold systems where the Coulomb and short-range interactions both act, the Coulomb interaction cannot be neglected. 
In the presence of the long-range Coulomb interaction, the scattering amplitude is known to exhibit qualitatively different behavior even at low energies, and the low-energy universality no longer appears~\cite{Bethe:1949yr,Domcke:1983zz,Kong:1998sx,Kong:1999sf,Ando:2007fh,Higa:2008dn,Mochizuki:2024dbf}. In this situation, it is natural to ask how the properties of near-threshold eigenstates differ from those of systems with purely short-range interactions.

In this contribution, we investigate the internal structure of near-threshold states in systems governed by the Coulomb and short-range interactions. Here, we focus on the case with an attractive Coulomb interaction. The case of a repulsive Coulomb interaction is discussed in Ref.~\cite{Kinugawa:2025kqr}.

\section{Formulation}

To study the eigenstates of systems with Coulomb and short-range interactions, we adopt the effective field theory developed in 
Refs.~\cite{Higa:2008dn}. The scattering amplitude is given as a function of the momentum $k$ and the scattering angle $\theta$:
\begin{align}
f(k,\theta) &= f_{C}(k,\theta) + f_{CS}(k,\theta), 
\label{eq:decomp}\\
f_{C}(k,\theta) &= -\frac{\eta^{2}}{2k_{C}}\frac{1}{\sin^{2}\left(\frac{\theta}{2}\right)}
\exp\left[2i\sigma_{0} - i\eta \log\left(\sin^{2}\frac{\theta}{2}\right) \right], 
\label{eq:fC-Gamma} \\
f_{CS}(k) &= C_{\eta}^{2}e^{2i\sigma_{0}}\left[-\frac{1}{a^{C}_{s}} + \frac{r^{C}_{e}}{2}k^{2} - 2k_{C}\left\{\psi(i\eta) + \frac{1}{2i\eta} - \log(-i\eta)\right\} \right]^{-1}.
\label{eq:fCS-renorm}
\end{align}
with $\eta = k_{C}/k$ and $k_{C} = -1/a_{B}$ for the attractive Coulomb 
interaction. The Coulomb phase shift for the $s$-wave 
scattering $\sigma_{0}$ and the Sommerfeld factor $C_{\eta}^{2}$ are defined as~\cite{Higa:2008dn}:
\begin{align}
\sigma_{0}=\frac{1}{2i}\log\frac{\Gamma(1 + i\eta)}{\Gamma(1 - i\eta)},\quad  
C_{\eta}^{2} &= \frac{2\pi\eta}{e^{2\pi \eta} - 1},
\label{eq:C-eta-2}
\end{align}
and $\Gamma(z)$ is the gamma function and $\psi(z)=(d/dz) \log\Gamma(z)$ is the digamma function. $a_{s}^{C}$ and $r_{e}^{C}$ are the Coulomb
scattering length and the Coulomb effective range, respectively. As in Eq.~\eqref{eq:decomp}, the scattering amplitude in the system with the short-range and Coulomb interactions are decomposed into two parts, the pure Coulomb scattering amplitude $f_{C}(k, \theta)$ in Eq.~\eqref{eq:fC-Gamma} and the Coulomb-distorted short-range amplitude $f_{CS}(k)$ in Eq.~\eqref{eq:fCS-renorm}~\cite{Higa:2008dn}. Here, $f_{CS}(k)$ is independent of $\theta$, since the contact short-range interaction without derivative is adopted.

The eigenmomentum is obtained from the pole condition of the Coulomb-distorted short-range amplitude~\eqref{eq:fCS-renorm}~\cite{Kong:1998sx,Kong:1999sf,Ando:2007fh,Higa:2008dn}:
\begin{align}
-\frac{1}{a_{s}^{C}} + \frac{r_{e}^{C}}{2}k^{2} - ik + \frac{2}{a_B}\left \{\log(-ia_Bk) + \psi\left(1 - \frac{i}{a_Bk}\right)\right\} &= 0.
\label{eq:pole-condition-ERE}
\end{align}
In the $a_{B} \to \infty$ limit, where the Coulomb interaction vanishes, Eq.~\eqref{eq:pole-condition-ERE} reduces to the effective range expansion (ERE) for a purely short-range interaction. This implies that the model adopted here can be regarded as an extension of the ERE that incorporates the Coulomb interaction. A solution of Eq.~\eqref{eq:pole-condition-ERE} on the positive imaginary $k$ axis, $k \equiv i\kappa$ with $\kappa > 0$ represents the bound state. On the other hand, a solution in the complex $k$ plane corresponds to the resonance. Due to the branch cut on the negative imaginary axis induced by the logarithmic term in the pole condition~\eqref{eq:pole-condition-ERE}, virtual states with $\kappa < 0$ do not appear in this system.

The compositeness $X$ is calculated by using the self-energy~\cite{Kinugawa:2025kqr}. The result can be expressed solely by 
observables, the effective range $r_{e}^{C}$ and the length scale $R^{C}$, which is determined by the eigenmomentum $k$ and the Bohr radius $a_{B}$:
\begin{align}
X &= \left[1 - \frac{r_{e}^{C}}{R^{C}}\right]^{-1}, 
\label{eq:wbr} \\
R^{C} &= -\frac{1}{k}\left[2i\left(\frac{1}{a_{B}k}\right)^{2}\psi_{1}\left(-i\frac{1}{a_{B}k}\right)+ i - 2\left(-\frac{1}{a_{B}k}\right) \right],
\label{eq:R-C}
\end{align}
where $\psi_{1}(z) = d\psi(z)/dz$ is the trigamma function. Recall that the compositeness $X$ of a shallow bound state formed by the short-range interaction can also be determined by the observables~\cite{Weinberg:1965zz}. The expression of $X$ in terms of the observables is known as the weak-binding relation~\cite{Kamiya:2016oao,Hyodo:2013nka,Kinugawa:2024crb}. In fact, in the limit $a_{B} \to \infty$ where the Coulomb interaction is switched off, we have $R_{c}\to R$ and $r_{e}^{C}\to r_{e}$ with the radius of the bound state wavefunction $R = 1/\sqrt{2\mu B}$ and the effective range $r_{e}$ by the purely short-range interaction. In this case, Eq.~\eqref{eq:wbr} reduces to the original weak-binding 
relation~\cite{Kamiya:2016oao}. 
Thus, Eq.~\eqref{eq:wbr} can be regarded as the extended weak-binding relation in the presence of the Coulomb interaction.

\section{Pole trajectories and compositeness}

\begin{figure}[tbp]
\centering
\includegraphics[width=0.45\textwidth]{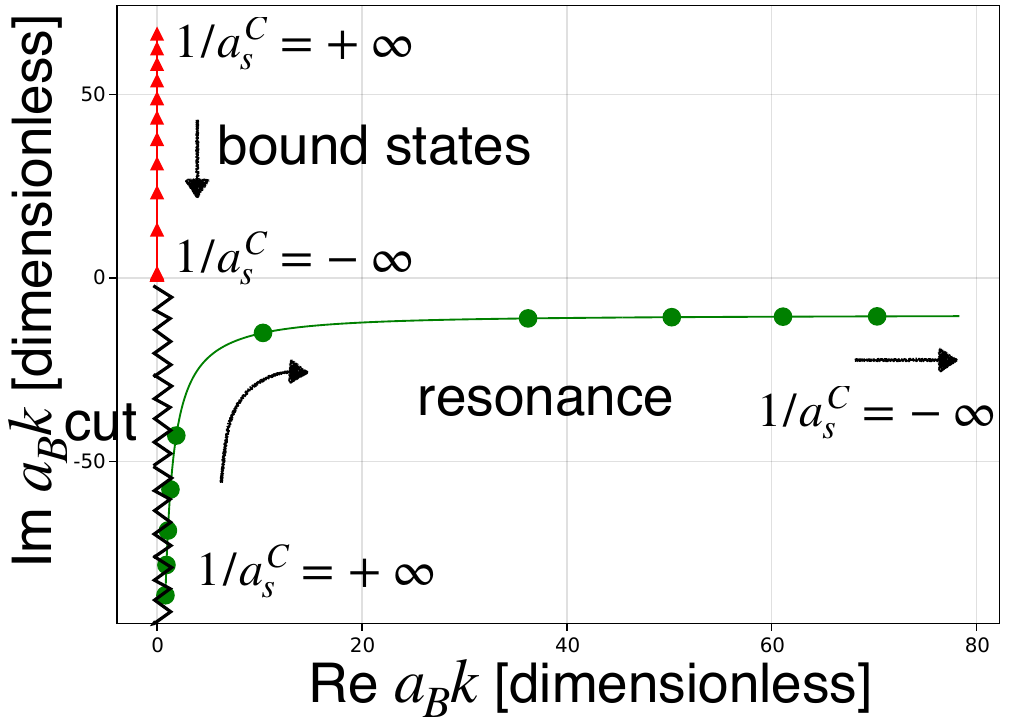}
\includegraphics[width=0.45\textwidth]{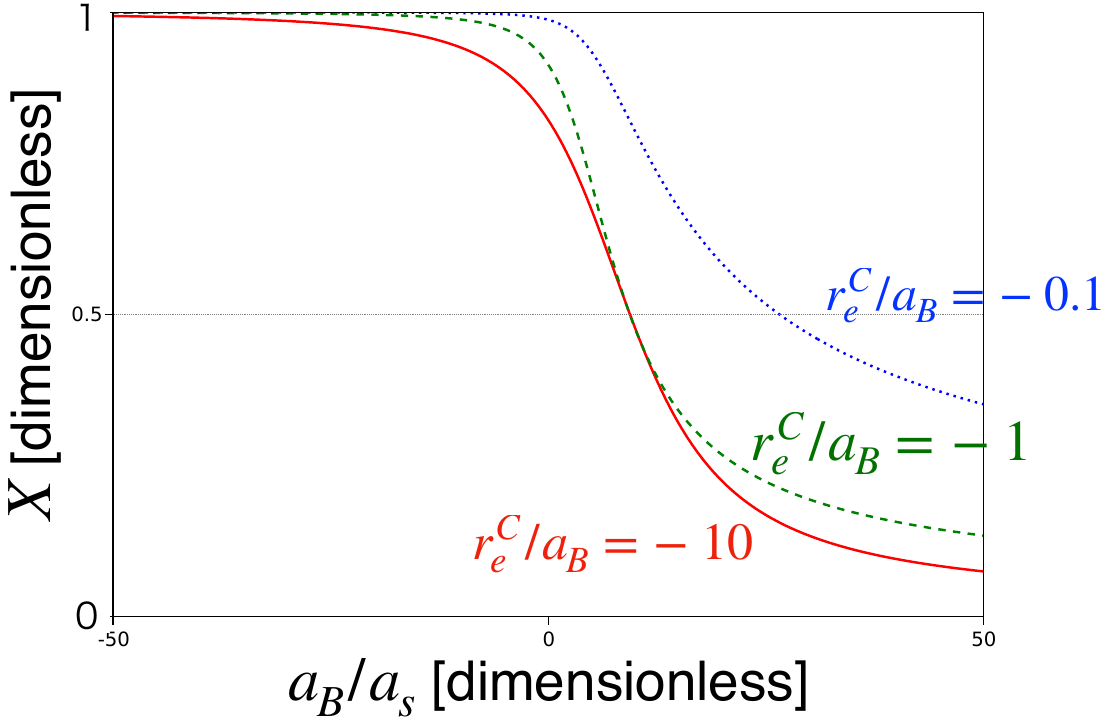}
\caption{Left panel: Pole trajectories of the eigenstates originated from the short-range interaction in the complex momentum $a_{B}k$ plane as the inverse scattering length $1/a_{s}^{C}$ is varied. The effective range is fixed at $r_{e}^{C}/a_{B} = -0.1$. Right panel: Compositeness $X$ of the bound state as functions of $a_{B}/a_{s}^{C}$ for $r_{e}^{C}/a_{B} = -10$ (solid line), $-1$ (dashed line), and $-0.1$ (dotted line). }
\label{fig:att}
\end{figure}

In this section, we present numerical results for systems with an attractive Coulomb interaction combined with short-range interactions. 
By solving the pole condition~\eqref{eq:pole-condition-ERE}, we compute the pole trajectories in the complex momentum plane under the variation of the inverse scattering length $1/a_{s}^{C}$, as shown in the left panel of Fig.~\ref{fig:att}. The effective range is fixed at $r_{e}^{C}/a_{B} = -0.1$. Due to the Schwarz reflection principle, the pole trajectories are symmetric with respect to the imaginary axis, so only the right half of the complex momentum plane is shown. In the presence of an attractive Coulomb interaction, there exist an infinite number of Coulomb originated bound states in the near-threshold region with ${\rm Im}\,a_{B} k \lesssim 1$. In this study, we focus on eigenstates that are formed mainly by the short-range interaction which are located outside the length scale set by the Bohr radius. For this reason, the Coulomb originated bound states are not shown in this figure. 

First, we focus on the trajectory of the bound state pole located on the positive imaginary axis. For a large $1/a_{s}^{C}\gg 1/a_{B}$, the pole is found far away from the origin, corresponding to a deeply bound state. As the inverse scattering length $1/a_{s}^{C}$ is decreased, the imaginary part of the eigenmomentum becomes smaller, indicating that the binding energy decreases. In the limit $1/a_{s}^{C} \to -\infty$, the pole approaches $a_{B}k=i$, the eigenmomentum of the ground state of the pure Coulomb bound states. 
In this way, the bound-state pole remains on the imaginary axis for any value of the scattering length, regardless of its sign. Note that as $1/a_{s}^{C}$ is varied from $-\infty$ to $+\infty$, each Coulomb-originated bound state is successively shifted to the next shallower level, as shown in Refs.~\cite{Domcke:1983zz,Mochizuki:2024dbf}.

Next, we discuss the pole located in the fourth quadrant of the $k$ plane. For a large $1/a_{s}^{C}$, the pole appears with a large negative imaginary part and a small real part. As the inverse scattering length $1/a_{s}^{C}$ is decreased, the pole moves toward the threshold with the real part remains almost unchanged. At around $1/a_{s}^{C}\sim 0$, the real part of the eigenmomentum increases rapidly, and eventually exceeds its imaginary part. Such state with positive excitation energy represents a resonance. As $1/a_{s}^{C}$ is further reduced, ${\rm Im}\, k$ appears to approach an almost constant value. We numerically find that the imaginary part of the resonance pole ${\rm Im}\, a_{B}k$ asymptotically approaches $a_{B}/r_{e}^{C} = -10$. It is shown that in the effective range expansion in the absence of the Coulomb interaction, the imaginary part of the pole of the resonance is given by the inverse of the effective range, ${\rm Im}\, k = 1/r_{e}$~\cite{Hyodo:2013iga}. This indicates that, in the limit $1/a_{s}^{C} \to -\infty$, the resonance pole in the presence of the Coulomb interaction asymptotically approaches that in the system with only short-range interactions.

We notice that, with the attractive Coulomb interaction, the pole trajectory of the bound state located on the positive imaginary axis does not continuously connected to that of the resonance in the fourth quadrant. This is in sharp contrast to the system with the repulsive Coulomb plus short-range interaction, where the bound state directly evolves into the resonance~\cite{Domcke:1983zz,Mochizuki:2024dbf,Kinugawa:2025kqr}, although both the systems have a bound state for a large positive $1/a_{s}^{C}$ and a resonance for a large negative $1/a_{s}^{C}$. With the attractive Coulomb interaction, there are infinitely many Coulomb-originated bound states near the origin, which prevent the short-range induced bound state pole to reach the origin due to the level repulsion. On the other hand, the scattering amplitude~\eqref{eq:fCS-renorm} has a resonance pole for $1/a_{s}^{C}\to -\infty$ with a fixed negative $r_{e}^{C}$, which should be supplemented by the pole originally exists in the fourth quadrant.

To quantitatively examine the internal structure of the bound state, we present the dependence of the compositeness $X$ on the inverse scattering length in units of the Bohr radius, $a_{B}/a_{s}^{C}$, in the right panel of Fig.~\ref{fig:att}. To controll the strength of the short-range interaction relative to the Coulomb force, we examine three cases, $r_{e}^{C}/a_{B} = -10$, $-1$, and $-0.1$, with larger magnitudes corresponding to weaker short-range interactions. As we discussed above, $a_{B}/a_{s}^{C}$ in the horizontal axis is related to the binding energy with large positive (negative) values corresponding to deeper (shallower) binding. In the large $a_{B}/a_{s}^{C}$ region, the compositeness is small in all cases, indicating that deeply bound states are non-composite dominant. As $a_{B}/a_{s}^{C}$ decreases, $X$ rapidly approaches unity, and for $1/a_{s}^{C} \lesssim 0$ the state becomes an almost purely composite state with $X \sim 1$. In fact, it can be analytically shown that $X$ become exactly unity in the limit $a_{B}/a_{s}^{C}\to -\infty$ where the pole represents the pure Coulomb bound state. By comparing results for different effective ranges, we observe that a smaller $|r_{e}^{C}|$ leads to a larger $X$ over the entire range of $a_{B}/a_{s}^{C}$ and to a faster convergence toward unity. This behavior can be understood as a remnant of the low-energy universality for purely short-range interactions, which becomes more pronounced for $|r_{e}^{C}/a_{B}|<1$, which is also observed in the system with repulsive Coulomb interaction~\cite{Kinugawa:2025kqr}. In this way, we find that near-threshold bound states in systems with the attractive Coulomb plus short-range interactions become almost purely composite as they sufficiently approach the threshold.
However, how the compositeness approaches unity is influenced by the remnant of the low-energy universality associated with the short-range interaction, depending on the strength of the short-range interaction relative to the Coulomb interaction.

\section{Summary}

In this contribution, we investigated the nature of near-threshold eigenstates in systems with the attractive Coulomb plus short-range interactions. Such systems were described using a model that reproduces the effective range expansion up to the effective range term, incorporating the contribution of the Coulomb interaction. The internal structure of near-threshold states was quantified by introducing the compositeness. We found that, in the presence of the attractive Coulomb interaction, the trajectories of bound-state and resonance poles are disconnected, which is qualitatively different from the results by the repulsive Coulomb plus short-range interaction. The compositeness of the deeply bound state is small but it approaches unity as the bound state comes close to the threshold. We showed that the manner in which the compositeness approaches unity is governed by the interplay between the Coulomb and short-range interactions, encoded in the Bohr raduis and the effective range, respectively.

\end{document}